# Jeans' criterion in nonextensive statistical mechanics

Du Jiulin

*Department of Physics, School of Science, Tianjin University, Tianjin 300072, China*
*E-mail*: jiulindu@yahoo.com.cn

**Abstract**

The Jeans' gravitational instability in nonextensive statistical mechanics is studied and a general form of the generalized Jeans' criterion is obtained that is related to the *q*-function $c_q = \sum_i p_i^q$. In this approach, the nonextensive model of classical ideal gas is applied to the Jeans' problem instead of the ordinary one in extensive statistical mechanics and the generalized critical wavelength to describe the gravitational instability is deduced. This nonextensive modification of the Jeans' criterion leads to a new critical length that depends not only on the nonextensive parameter *q* but also on the dimension *D* and the total particle numbers *N* of the system. When $q \to 1-$, the Jeans' length is perfectly recovered. We also give the nonextensive parameter *q* a physical interpretation that $q = 1$ represents an isothermal process of the gas, corresponding to the state of complete mixing, but 0< *q* <1 is nonisothermal, corresponding to the state of incomplete mixing, it measures the degree of mixing.

## 1. Introduction

Many-body gravitating systems play a very essential role in astrophysics and cosmology. The dynamical stability of a self-gravitating system usually can be described by the Jeans' criterion for the gravitational instability. The equations of the Jeans' problem are the equation of continuity, the Euler's equation, the Poisson's equation and the equation of state of an ideal gas (in the isothermal form)[1]:

$$\frac{\partial \rho}{\partial t} + \nabla(\rho \mathbf{v}) = 0 \qquad (1)$$

$$\frac{\partial \mathbf{v}}{\partial t} + (\mathbf{v} \cdot \nabla)\mathbf{v} = -\frac{1}{\rho}\nabla P - \nabla \varphi \qquad (2)$$

$$\nabla^2 \varphi = 4\pi G \rho \qquad (3)$$

$$P = \frac{kT}{\mu \, m_H} \rho, \quad (T \text{ is constant}) \tag{4}$$

where, as usual, $\rho$ is the mass density, **v** the velocity, $P$ pressure, $T$ temperature, $G$ gravitational constant, $\mu$ the mean molecular weight, $m_H$ the atomic mass of hydrogen, and $\varphi$ the gravitational potential. On the basis of these equations, the conclusion of Jeans' criterion of gravitational instability is that density fluctuations with the wavelength $\lambda$ more than the critical value $\lambda_J = v_s \sqrt{\pi / G \rho_0}$ will grow so that the system becomes gravitationally unstable [1]. This means that an isothermal gaseous sphere with the length scale more than $\lambda_J$ is gravitational instability and is going to contract constantly. In this formula, $\lambda_J$ is called Jeans' length, $v_s = \sqrt{kT / \mu \, m_H}$ is the sound speed and $\rho_0$ is the undisturbed mass density.

On the other hand, the statistical mechanics of self-gravitating systems with long-range interaction exhibits the peculiar features such as negative specific heats [2-5] and the so-called gravithermal instability [6,7], which are greatly different from the usual thermodynamic systems and are therefore quite difficult to understand. The structure and stability of self-gravitating systems at statistical equilibrium are analyzed usually in terms of the maximization of a thermodynamic potential (the so-called mean field description). This thermodynamic approach leads to isothermal configurations that have been studied for long time in the stellar structure and the galactic structure [8-11]. It is well known that isothermal configurations only correspond to meta-stable states, not true equilibrium states.

With the above phenomenon, it has been considered that systems with the gravitational long-range interaction may be nonextensive and, therefore, the conventional Boltzmann-Gibbs statistical mechanics may be not appropriate to the description of the features of the systems. In recent years, a nonextensive generalization of Boltzmann-Gibbs statistical mechanics has focused significant attention, which has been known as "Tsallis statistics"[12].

Tsallis made the nonextensive generalization of Boltzmann-Gibbs entropy by constructing the following form of entropy [13]:

$$S_q = \frac{k}{1-q}\left(\sum_i p_i^q - 1\right) \tag{5}$$

where $k$ is the Boltzmann constant, $p_i$ the probability that the system under consideration is in its $i$th configuration such that $\sum_i p_i = 1$, and $q$ a positive nonextensive parameter whose deviation from unity describes the degree of nonextensivity of the system. The celebrated Boltzmann entropy $S_B$ is so obtained as the limit $q \to 1$:

$$S_B = \lim_{q \to 1} S_q = -k \sum_i p_i \ln p_i \tag{6}$$

The fundamental difference between the Tsallis entropy $S_q$ and the Boltzmann entropy $S_B$ lies in that if the system is composed of two independent subsystems, $a$ and $b$, then the total Tsallis entropy of the system satisfies $S_q(a+b) = S_q(a) + S_q(b) + (1-q)S_q(a)S_q(b)/k$. The recent developments of this theory are now referred to as Nonextensive Statistical Mechanics (NSM).

In this paper, we apply NSM to the study of Jeans' gravitational instability of a self-gravitating system. We use a nonextensive form of the state equation of an ideal gas to the Jeans' problem and deduce a generalized analytical expression of the Jeans' criterion of gravitational instability, in which we express a nonextensive modification for the critical length. The well-known Jeans' length is perfectly recovered when $q \to 1$. We also give a physical interpretation about the nonextensive parameter $q$ in the final section.

## 2. Some results of NSM

In NSM, the probability distribution function [14,15] is

$$p_i = \frac{1}{Z_q}\left[1 - (1-q)\frac{\beta(\varepsilon_i - U_q)}{c_q}\right]^{1/(1-q)} \tag{7}$$

where $Z_q$ is the generalized canonical partition function defined [15] by

$$Z_q = \sum_i \left[1 - (1-q)\frac{\beta(\varepsilon_i - U_q)}{c_q}\right]^{1/(1-q)} \tag{8}$$

$c_q = \sum_i p_i^q$, $U_q = \frac{1}{c_q}\sum_i \varepsilon_i p_i^q$ is the generalized internal energy, and $\beta$ is the Lagrange multiplier associated with this energy constraint. It has been proved in NSM that $\beta$ is identified with the usual inverse temperature [14,15] defined as

$$k\beta = \frac{\partial S_q}{\partial U_q} = \frac{1}{T} \tag{9}$$

From Eqs.(5), (7) and (8) one finds

$$c_q = \sum_i p_i^q = Z_q^{1-q} \tag{10}$$

$$S_q = \frac{k}{1-q}(c_q - 1) = \frac{k}{1-q}(Z_q^{1-q} - 1) \tag{11}$$

According to the generalized zeroth law of thermodynamics, the physical temperature $T_q$ and the physical pressure $P_q$ in NSM are variables and are defined [16,19] by

$$T_q = (1 + \frac{1-q}{k}S_q)\left(\frac{\partial S_q}{\partial U_q}\right)^{-1} \tag{12}$$

$$P_q = \frac{T_q}{1+(1-q)S_q/k}\left(\frac{\partial S_q}{\partial V}\right) \tag{13}$$

where $V$ is volume of the system. Substituting Eqs.(9), (10) and (11) into these two equations one obtains

$$T_q = (1 + \frac{1-q}{k}S_q)T = c_q T \tag{14}$$

$$P_q = T_q \frac{k}{1-q}\frac{\partial}{\partial V}\ln c_q \tag{15}$$

or $\quad P_q = kT_q \frac{\partial}{\partial V}\ln Z_q \tag{16}$

Thus the physical temperature $T_q$ in NSM is different from the inverse of the Lagrange multiplier $\beta$ and so different from the temperature $T = 1/k\beta$ in extensive systems. The physical pressure $P_q$ is also not the ordinary one. However, It is readily proved that the equation of state for a classical ideal gas in NSM has the same form as that in ordinary extensive statistical mechanics.

The generalized partition function of a classical ideal gas [16] is

$$Z_q = \frac{\Gamma\left(\frac{2-q}{1-q}\right)}{\Gamma\left(\frac{2-q}{1-q}+\frac{DN}{2}\right)} \frac{V^N}{N!h^{DN}} \left(\frac{2\pi\, mkT_q}{1-q}\right)^{\frac{DN}{2}} \left[1+(1-q)\frac{DN}{2}\right]^{\frac{1}{1-q}+\frac{DN}{2}} \tag{17}$$

Substituting into Eq.(16) one easily obtains

$$P_q V = NkT_q \tag{18}$$

By taking the particle mass $m = \mu\, m_H$ and the particle number density $n = N/V = \rho/\mu\, m_H$ into account, one can write Eq.(18) in the form

$$P_q = \frac{k}{\mu\, m_H} \rho\, T_q \tag{19}$$

## 3. Jeans' criterion in NSM

In NSM, the Jeans' problem of a self-gravitating system should be described by Eqs.(1), (2), (3) and (19). Let $\rho_1$, $\mathbf{v}_1$, $\varphi_1$ and $P_{q1}$ be small perturbations in the gas about its equilibrium state $\rho_0$, $\mathbf{v}_0$, $\varphi_0$ and $P_{q0}$, respectively. Assume the conditions under which the system is static and homogeneous when it is in its equilibrium to be that $\mathbf{v}_0 = 0$, $\rho_0$ and $P_{q0}$ are constant, respectively, then, with $\rho = \rho_0 + \rho_1$, $\mathbf{v} = \mathbf{v}_0 + \mathbf{v}_1$, $\varphi = \varphi_0 + \varphi_1$ and $P_q = P_{q0} + P_{q1}$ one obtains the linearized equations

$$\frac{\partial \rho_1}{\partial t} + \rho_0 \nabla \cdot \mathbf{v}_1 = 0 \tag{20}$$

$$\frac{\partial \mathbf{v}_1}{\partial t} = -\nabla \varphi_1 - \frac{1}{\rho_0} \nabla P_{q1} \tag{21}$$

Considering

$$T_q = T_q(\rho) = T_q(\rho_0 + \rho_1) = T_q(\rho_0) + T_q'(\rho_0)\rho_1 + \tfrac{1}{2} T_q''(\rho_0)\, \rho_1^2 + \cdots\cdots,$$

and using Eq.(19), one has $P_{q0} = \frac{k}{\mu m_H} \rho_0 T_q(\rho_0)$ and

$$P_{q1} = \frac{k}{\mu m_H} [\rho_0 T_q'(\rho_0) + T_q(\rho_0)] \rho_1 \tag{22}$$

Because of $T_q(\rho) = c_q(\rho)T$, Eq.(22) becomes

$$P_{q1} = \frac{kT}{\mu m_H} C_q(\rho_0) \rho_1 \tag{23}$$

where $C_q(\rho_0) = \rho_0 c_q'(\rho_0) + c_q(\rho_0)$. Substituting Eq.(23) into Eq.(21) one has

$$\frac{\partial \mathbf{v}_1}{\partial t} = -\nabla \varphi_1 - \frac{1}{\rho_0} \frac{kT}{\mu m_H} C_q(\rho_0) \nabla \rho_1 \tag{24}$$

The amalgamation of Eq.(24) and Eq.(20) and the use of Eq.(3) yield

$$\frac{\partial^2 \rho}{\partial t^2} = 4\pi G \rho_0 \rho_1 + \frac{kT}{\mu m_H} C_q(\rho_0) \nabla^2 \rho_1 \tag{25}$$

Taking the perturbation in the form of $\rho_1 \sim \exp\left[\omega t + i(\frac{2\pi x}{\lambda})\right]$, then one finds the dispersion relation:

$$\omega^2 = 4\pi G \rho_0 - \left(\frac{2\pi}{\lambda}\right)^2 \frac{kT}{\mu m_H} C_q(\rho_0) \tag{26}$$

From this equation one can obtain the critical wavelength $\lambda_C$

$$\lambda_C = \lambda_J \sqrt{C_q(\rho_0)} \tag{27}$$

where $\lambda_J = v_s \sqrt{\pi / G \rho_0}$ is the Jeans' length and $v_s = \sqrt{kT / \mu m_H}$ the sound speed [1]. Eq.(27) gives the conclusion that if the wavelength $\lambda$ of density fluctuation is more than the critical one $\lambda_C$, the density will grow with time in the exponent form, and the system will become gravitationally unstable. Thus, in NSM, the Jeans' criterion for the gravitational instability is modified as

$$\lambda > \lambda_J \sqrt{C_q(\rho_0)} \tag{28}$$

which is related to the $q$-function $c_q = \sum_i p_i^q$ through the factor $\sqrt{C_q(\rho_0)}$. It is easily proved that when $q \to 1$, $C_q \to 1$ and this generalized Jeans' criterion is recovered to the familiar form perfectly.

## 4. A nonextensive model of classical ideal gas

The Hamiltonian of a classical ideal gas in $D$-dimension space reads $H = \sum_i^N \mathbf{p}_i^2 / 2m$, where $\mathbf{p}_i$ is the $D$-dimension momentum of the $i$th particle ( $i$ =1, 2, …… $N$ ), m is the particle mass and $N$ is the total particle numbers. In NSM, the probability distribution function of the system is determined by the maximization of Tsallis entropy [15-18] as

$$f(\{\mathbf{p}_i\}) = \frac{1}{Z_q} \left[ 1 - (1-q) \frac{\beta}{c_q} \left( \sum_i^N \frac{\mathbf{p}_i^2}{2m} - U_q \right) \right]^{\frac{1}{1-q}} \tag{29}$$

which defines the undisturbed states of the system with the long-range interaction, where

$$c_q = \left[ \frac{\Gamma[1/(1-q)]}{\Gamma[1/(1-q) + DN/2]} \frac{V^N}{N! h^{DN}} \left( \frac{2\pi mkT}{1-q} \right)^{\frac{DN}{2}} \right]^{\frac{2(1-q)}{2-(1-q)DN}} \times \left[ 1 + (1-q) \frac{DN}{2} \right]^{\frac{2q+(1-q)DN}{2-(1-q)DN}} \tag{30}$$

and 0< $q$ <1. From Eq.(14) and (18), straightforwardly one has

$$P_q = \frac{NkT}{V} c_q \tag{31}$$

When $q \to 1$, $c_q \to 1$ and, therefore, the ordinary equation of state of an ideal gas is recovered perfectly from the above equation. Using the Stirling's formula in Eq.(30), for a large $N$ one finds

$$\frac{V^N}{N!} \approx \frac{e^N}{\sqrt{2\pi N}} m^{-N} \rho^{-N} \tag{32}$$

Eq.(30) is therefore written as

$$c_q(\rho) = A_q(D, N) \rho^{\frac{2N(1-q)}{DN(1-q)-2}} \tag{33}$$

with the factor $A_q(D, N)$:

$$A_q = \left[ \frac{\Gamma[1/(1-q)]}{\Gamma[1/(1-q) + DN/2]} \frac{e^N m^{(\frac{D}{2}-1)N}}{h^{DN} \sqrt{2\pi N}} \left( \frac{2\pi kT}{1-q} \right)^{\frac{DN}{2}} \right]^{\frac{2(1-q)}{2-(1-q)DN}} \left[ 1 + (1-q)\frac{DN}{2} \right]^{\frac{2q+(1-q)DN}{2-(1-q)DN}}$$

where m=$\mu m_H$. Substituting Eq.(33) into Eq.(27), one obtains the critical wavelength $\lambda_C$ on the gravitational instability

$$\lambda_C = \lambda_J \left[ \frac{2N(1-q)}{DN(1-q)-2} + 1 \right]^{\frac{1}{2}} [c_q(\rho_0)]^{\frac{1}{2}} \tag{34}$$

It is clear that when $q \to 1-$ the Jeans' length $\lambda_J$ of extensive system is correctly recovered from Eq.(34). This new formula indicates that $\lambda_C$ not only depends on $q$ but on $D$ and $N$ of the system.

## 5. Conclusions

In conclusion, we have studied the Jeans' problem in NSM and have obtained the new generalized Jeans' criterion given by (28). The criterion (28) is a general form of the generalized Jeans' criterion in NSM, which is related to the $q$-function $c_q = \sum_i p_i^q$ of a nonextensive system and so the specific form of which can be determined by the nonextensive model of system. In our approach, the nonextensive model of a classical ideal gas is applied to the analyses of Jeans' gravitational instability instead of the ordinary one in extensive statistical mechanics. In this case, the physical temperature is no longer constant. From hydrodynamic stability analyses of the problem, we deduce the generalized critical wavelength that describes the gravitational instability of a self-gravitating system. This nonextensive modification of the Jeans' criterion leads to a new critical length that depends not only on the nonextensive parameter $q$ but also on the dimension $D$ of the space and the total particle numbers $N$ of the system. When $q \to 1-$, the Jeans' length in extensive systems is correctly recovered.

In this way, we can give the nonextensive parameter $q$ a physical interpretation. $q = 1$ represents an isothermal processes of the gas, corresponding to the state of complete mixing, but 0< *q* <1 is nonisothermal, corresponding to the state of incomplete mixing, it measures the degree of mixing. However, as talked in the introduction, the isothermal state only corresponds to a meta-stable state (locally mixing) of a self-gravitating system, not the true equilibrium state. In view of this point, our results make it clear that NSM may be the appropriate approach for the description of physical features of self-gravitating systems.

Recent works [20] on the Jeans' gravitational instability in nonextensive kinetics have attracted our attentions. With a *phenomenological* nonextensive generalization of the Maxwellian distribution function [21], these authors studied nonextensive modification of the Jeans' criterion. In fact, in the view of kinetics Ref.[21] provided a nonextensive gas model. With this model one finds the mean square velocity of the particle [22]

$$<v^2> = \int v^2 f(v) d^3v = \frac{6}{3q-1}\frac{kT}{m}, \quad (q > \frac{1}{3}). \tag{35}$$

and the nonextensive equation of state of the gas

$$P_q = \frac{1}{3}nm<v^2> = \frac{2}{3q-1}nkT \tag{36}$$

If one solves the Jeans problem that is determined by Eqs.(1), (2),(3) and (36), or if one compares Eq.(36) with Eq.(18) and substitute $c_q = 2/(3q-1)$ into the criterion (28), one can immediately obtain the results of Ref.[20] concerning the Jeans' criterion.

**Acknowledgments**

The author would like to thank Professor S. Abe for his useful discussion of the nonextensive model of classical ideal gas, Professor R. Silva for the useful discussion of the *q*-nonextensive velocity distribution and Professor J. A. S. Lima for his recent reprint. This work is supported by the project of the "985" Program of TJU.

# References

[1] K.R.Lang, Astrophysical Formulae, 1974, Springer-Verlage.

[2] A.S.Eddington, The Internal Constitution of Stars, 1926, Cambridge University Press.

[3] D.Lynden-Bell, R.Wood, MNRAS, **138**(1968)495.

[4] S.W.Hawking, Nature, **248**(1974)30.

[5] T.Padmanabhan, Phys.Rep., **188**(1990)285.

[6] A.Taruya, M.Sakagami, Physica A, **307**(2002)185.

[7] A.Taruya, M.Sakagami, Physica A, **318**(2003)387.

[8] S.Chandrasekhar, An Introduction to the Theory of Stellar Structure, 1942, Dover.

[9] J.Binney, S.Tremaine, Galactic Dynamics, 1987, Princeton Series in Astrophysics.

[10] P.H.Chavanis, Astron.Astrophys., **401**(2003)15.

[11] P.H.Chavanis, Astron.Astrophys., **381**(2002)340.

[12] S.Abe, A.K.Rajagopal, Science, **300**(2003)249.

[13] C.Tsallis, J.Stat.Phys., **52**(1988)479.

[14] C.Tsallis, R.S.Mendes, A.R.Plastino, Physica A, **261**(1998)534.

[15] S.Abe, Phys.Letts.A, **263**(1999)424; **267**(2000)456.

[16] S.Abe, S.Martinez, F.Pennini, A.Plastino, Phys.Letts.A, **281**(2001)126.

[17] S.Abe, S.Martinez, F.Pennini, A.Plastino, Phys.Letts.A, **278**(2001)249.

[18] S.Abe, Physica A, **269**(1999)403

[19] R.Toral, Physica A, **317**(2003)209.

[20] J.A.S.Lima, R.Silva, J.Santos, Astron.Astrophys., **396**(2002)309.

[21] R.Silva, A.R.Plastino, J.A.S.Lima, Phys.Letts.A, **249**(1998)401.

[22] J.A.S.Lima, R.Silva, J.Santos, Phys. Rev.E, **61**(2000)3260.